\documentclass[onecolumn]{aa}
\usepackage{graphicx}
\usepackage{natbib}

\begin{document}
\title{Neutrino emission rates in highly magnetized neutron stars revisited}

\author{Mario Riquelme\inst{1}, Andreas Reisenegger\inst{1},
Olivier Espinosa\inst{2}, \& Claudio O. Dib\inst{2}}
\institute{Departamento de Astronom\'\i a y Astrof\'\i sica,
Pontificia Universidad Cat\'olica de Chile, Casilla 306, Santiago
22, Chile. \and Departamento de F\'\i sica, Universidad T\'ecnica
Federico Santa Mar\'\i a, Casilla 110-V, Valpara\'\i so, Chile.}

\date{Received / Accepted}

\abstract{ Magnetars are a subclass of neutron stars whose intense
soft-gamma-ray bursts and quiescent X-ray emission are believed to
be powered by the decay of a strong internal magnetic field.
We reanalyze neutrino emission in such stars in the plausibly
relevant regime in which the Landau band spacing $\Delta
E$ of both protons and electrons is much larger than $kT$
(where $k$ is the Boltzmann constant and $T$ is the temperature),
but still much smaller than the Fermi energies. Focusing on the
 direct Urca process, we find that the emissivity oscillates as a function of density or
magnetic field, peaking when the Fermi level of the protons or
electrons lies about $\sim 3 kT$ above the bottom of any of their
Landau bands. The oscillation amplitude is comparable to the
average emissivity when $\Delta E$ is roughly the geometric mean
of $kT$ and the Fermi energy (excluding mass), i. e., at fields
much weaker than required to confine all particles to the lowest
Landau band. Since the density and magnetic field strength vary
continuously inside the neutron star, there will be alternating
surfaces of high and low emissivity. Globally, these oscillations
tend to average out, making it unclear whether there will be any
observable effects.}


\titlerunning{Neutrino emission in magnetized neutron stars}
\authorrunning{M. Riquelme et al.}
\maketitle


\section{Introduction}

\label{sec:in}

In recent years, evidence has been accumulating that soft
gamma-ray repeaters and anomalous X-ray pulsars are {\it
magnetars}, strongly magnetized neutron stars powered by the decay
of their magnetic field (see Woods \& Thompson 2004 for a recent
review).
The simplest estimate of their magnetic field strength relies on
the magnetic dipole braking model usually applied to radio
pulsars, which for these objects yields a very strong dipole
field, $\sim 10^{14-15}$ G (Kouveliotou et al. 1998; Hurley et al.
1999). Contrary to the case of pulsars, their time-averaged
luminosity (in the form of quiescent X-rays and soft gamma-ray
bursts) by far exceeds their rotational energy loss, demanding
another source of energy, which might be the decay of their
internal magnetic field (Duncan \& Thompson 1992; Paczy\'nski
1992; Thompson \& Duncan 1995, 1996), if this is substantially
stronger than the external dipole field.

The observable properties of these objects depend, among
other things, on their field strength and their temperature, whose
evolution is coupled through the energy dissipated by the field
decay and through the thermal modulation of the decay rates
(Thompson \& Duncan 1996). The most effective cooling mechanism
for hot neutron stars is neutrino emission (see, e.g., Yakovlev \&
Pethick 2004), which also may to some extent control the magnetic
field decay (Pethick 1992; Goldreich \& Reisenegger 1992).
Particularly effective, if allowed,
are the direct Urca processes, $n \rightarrow p + e^{-} +
\overline{\nu}_e$ and $p + e^{-} \rightarrow n + \nu_e$, involving
neutrons ($n$), protons ($p$), electrons ($e^{-}$), electron
neutrinos ($\nu_e$), and electron antineutrinos
($\overline\nu_e$). However, the matter inside a neutron star is
highly degenerate. In the absence of an extremely strong magnetic
field (far stronger than those discussed above; see Baiko \&
Yakovlev 1999), Pauli blocking, together with energy and momentum
conservation, requires the triangle inequalities

\begin{equation}
|p_{Fp} - p_{Fe}| \leq p_{Fn} \leq p_{Fp} + p_{Fe}
\label{eq:condition}
\end{equation}

\noindent to be at least approximately satisfied for the direct
Urca processes to be possible ($p_{Fi}$ is the Fermi momentum of
particle $i$, calculated in the non-magnetic approximation).
Still, large uncertainties in the composition and equation of
state of matter inside neutron stars cores prevent us from
deciding whether this condition is fulfilled (e. g., Yakovlev \&
Pethick 2004). In this paper, we assume that it is, and study the
effects of a magnetar-strength field on this simplest neutrino
emission process by accounting explicitly for the Landau-band
structure of the proton and electron energy levels. We focus on
the direct neutron $\beta$ decay ($n \rightarrow p + e^{-} +
\overline{\nu}_e$) in chemical equilibrium. Other reactions should
be affected in a qualitatively similar way, including modified
Urca processes in chemical equilibrium or any reactions away from
equilibrium, responsible for bulk viscosity (e. g., Finzi 1965;
Haensel 1992; Reisenegger 1995).

The magnetic field can make the emissivity change by two effects (Baiko $\&$ Yakovlev 1999).
The first relates to the non-trivial spatial form of the wave
functions of the charged particles. The second is an increased
contribution to the emissivity from the highest Landau band
occupied by electrons and/or protons. If the density or the
magnetic field are changed, the emissivity oscillates (with an amplitude
controlled by the temperature) as new Landau bands become
populated or depopulated. The purpose of this
work is to perform a detailed study of the second of these effects,
quantifying its importance as a function of the different
physical parameters involved.

In order to justify the assumptions made in our calculations and
to specify the situations in which this work applies, in
\S~\ref{sec:pc} we characterize the physical conditions we assume
inside the star. In \S~\ref{sec:er}  we describe the calculations
performed and the results obtained. A general discussion of the
relevance of the results is given in \S~\ref{sec:discussion},
whereas \S~\ref{sec:co} lists our main conclusions.

\section{Physical Conditions}
\label{sec:pc} It is known that a magnetic field $B$ modifies the
energy eigenstates and eigenvalues of electrons, protons and
neutrons, whose masses we denote as $m_e$, $m_p$, and $m_n$,
respectively. The energies of free electrons and protons are given
by
\begin{equation}
E_e = \sqrt{m^2_ec^4 + p^2_{ez}c^2 + 2Be\hbar cn}\quad\textrm{
and }\quad E_p = \sqrt{m^2_pc^4 + p^2_{pz}c^2 + 2Be\hbar cm},
\label{eq:energies}
\end{equation}
where $c$ is the speed of light, $e$ is the charge of the proton,
$\hbar$ is the (reduced) Planck constant, and
\begin{equation}
m = l + \frac{1}{2} - \frac{\sigma_p}{4}g_p, \quad {\rm
where}\quad l=0,1,2,\ldots\quad{\rm and}\quad \sigma_p = \pm 1.
\label{eq:proton}
\end{equation}
The integer $n$ and non-integer $m$ label the {\it Landau bands}
of electrons\footnote{Note that, except for the $n=0$ band, all
the Landau bands of electrons are spin-degenerate, when QED corrections to $g_e=2$ are neglected.} and protons,
respectively, $\sigma_p$ is the doubled proton spin component
along the direction of the magnetic field, $g_p=2.79$ is the
gyromagnetic ratio of the proton, and $p_{iz}$ is the component of
the linear momentum of the particle $i$ along the magnetic field
direction.

We note that, within each Landau band, the energy increases
continuously with the absolute value of the longitudinal momentum
component, $|p_{iz}|$, starting from a bottom value corresponding
to $p_{iz}=0$, which, for simplicity, we will refer to as the {\it
Landau level} corresponding to this band. As the density increases
in degenerate matter at a given magnetic field strength, new
Landau bands get populated as the Fermi energy, $E_{Fi}$,
increases, leading to oscillating thermodynamic functions (see, e.
g., Dib \& Espinosa 2001). Here, our main focus will be on similar
oscillations in the neutrino emissivity.

We ignore the effect of inter-particle forces on the energy
levels of the charged particles. This is a good approximation for the
electrons, which do not interact strongly. However, strong
interactions do modify the eigenvalues of energy  of the protons,
whose Landau levels tend to broaden. We discuss below the importance of this
broadening for the contribution to the emissivity from the highest
Landau band occupied by protons.

If we neglect strong interactions for the
neutrons as well, the second triangle inequality in eq.
(\ref{eq:condition}) would never be satisfied, and direct Urca
reactions would be forbidden (e.g., Shapiro \& Teukolsky 1983).
Thus, for consistency, we allow for an arbitrary dispersion
relation $E_n(p_n)$ for the neutrons, which takes their strong
interactions into account and could in principle allow direct Urca
processes to take place, even in the absence of a magnetic field.
This relation defines an effective mass for the neutron,
$m_n^*\equiv p_{Fn}/[dE_n/dp_n]_{p_{Fn}}$, which is generally
somewhat smaller than the ``bare'' mass, $m_n$. Here and
throughout this paper, we ignore the relatively small dependence
of the neutron dispersion relation on the spin projection along
the magnetic field.

In order to specify the physical conditions to be considered, we
define the dimensionless quantities

\begin{equation}
n_M \equiv \frac{E^2_{Fe} - m_e^2c^4}{2Be\hbar c} \textrm{ , } m_M
\equiv \frac{E^2_{Fp} - m_p^2c^4}{2Be\hbar c} ,  \textrm{ and } r
\equiv \frac{p_{Fn}^2c}{2Be\hbar} , \label{eq:def1}
\end{equation}

\noindent where the integer part of $n_M$ corresponds to the
highest Landau band populated by electrons in the zero-temperature
limit. The same role is played for the protons by the largest
value of $m\leq m_M$ compatible with eq. (\ref{eq:proton}). We
note that each of these can quantitatively be written as $\approx
150\rho_{37}^{2/3}/B_{16}$, where the number density of the
relevant particles is $\rho=\rho_{37}\times 10^{37}{\rm cm}^{-3}$
and the magnetic field is $B=B_{16}\times 10^{16}{\rm G}$. It is
also useful to identify the energy difference between adjacent
Landau levels for the particle $i=e,p$,

\begin{equation}
\Delta E_{i} \sim \frac{Be\hbar c}{E_{i}}, \label{eq:def2}
\end{equation}

\noindent which takes the values $\Delta E_e\approx
0.5B_{16}/\rho_{37}^{1/3}$ MeV for the (highly relativistic)
electrons and $\Delta E_p\approx 0.06B_{16}$ MeV for the
(approximately non-relativistic) protons. We assume that the
electrons, protons, and neutrons are in chemical equilibrium and
consider, for the calculation of the transition probabilities of
the direct $\beta$ decay, that the protons and neutrons are
non-relativistic particles.

Most importantly, we will see below that, in order for
the Landau band structure to be noticeable, the magnetic field
must be at least strong enough to make $\Delta E_{i} \sim 10kT$,
where $k$ is the Boltzmann constant and $T$ is the temperature.
However, magnetic fields expected to exist in magnetars are still
so weak as to make $r$, $n_M$, and $m_M \gg 1$, which means that
the electrons and protons populate many Landau bands. Writing
$T=T_8\times 10^8{\rm K}$, these conditions constrain the magnetic
field to the range
\begin{equation}
T_8\ll B_{16}\ll 10^2\rho_{37}^{2/3}. \label{eq:range}
\end{equation}

\noindent While the upper bound is easily satisfied by all observational
evidence, the lower bound is not obviously true in any known case.
Indeed, the lower bound (here given for protons, although a less
stringent condition for electrons also exists: $0.1
T_8\rho_{37}^{1/3} \ll B_{16}$) might not be satisfied even for
magnetars, despite their very strong $B$ fields, because the
dissipation of their huge magnetic energy tends to keep their
interiors fairly hot. If that is the case, the magnetic field
effects we are studying here are negligible. Nevertheless, given
the possibility that this regime may be applicable, we proceed to
explore its consequences.


\section{Emission rate calculations}
\label{sec:er}

We calculate the neutrino emissivity due to the direct $\beta$
decay using the Weinberg-Salam-Glashow theory of weak
interactions. We confirm Baiko \& Yakovlev's (1999) eqs. (5) and
(8) for the power radiated per unit volume,
\begin{equation}
Q_{\bar\nu}=\frac{2Be}{(2\pi)^7\hbar^9c}
\sum_{n,l,\sigma_p,\sigma_n}\int d^3p_n d^3p_{\bar\nu} dp_{ez}
dp_{pz} E_{\bar\nu} f_n(1-f_p)(1-f_e)\delta (\Delta E)
\delta(\Delta p_z)M, \label{eq:gf}
\end{equation}
where
\begin{equation}
M=G_F^2\left\{2g_A^2\left[\delta_{\sigma_p,1}\delta_{\sigma_n,-1}
F_{n,l}^2(u)+\delta_{\sigma_p,-1}\delta_{\sigma_n,1}
F_{n-1,l}^2(u)\right] + {1\over 2}\delta_{\sigma_p,\sigma_n}
\left[(1+g_A\sigma_p)^2F_{n-1,l}^2(u)+(1-g_A\sigma_p)^2F_{n,l}^2(u)\right]\right\},
\label{eq:matrix}
\end{equation}
$u=c\{p_n^2 - (p_{ez} + p_{pz})^2\}/(2Be\hbar)$,
$\Delta E = E_n-E_p-E_e-E_{\bar{\nu}}$, $\Delta
p_z=p_{nz}-p_{ez}-p_{pz}$, and $f_i$ is the Fermi-Dirac factor
corresponding to the particle $i$. The Fermi weak coupling
constant is $G_F = 1.399 \times 10^{-49}$ $\textrm{erg} \textrm{
}\textrm{cm}^3$, and $g_A = 1.261$ is the nucleon axial-vector
coupling. The Laguerre functions $F_{n,l}(u)$ are defined in the
Appendix (and are taken to vanish when one of their indices is
negative).
The physical conditions discussed in \S \ref{sec:pc} have already
been used to neglect the antineutrino momentum, $p_{\bar\nu}\sim
kT/c\ll p_{Fn}$, both in the definition of the variable $u$ and in
the momentum-conserving delta function. In principle, it
contributes to the ``thermal smoothing'' discussed below, but to a
lesser degree than the antineutrino energy, which we keep in our
calculation.

The integrals over directions of motion for antineutrinos and
neutrons are easily done (the latter eliminates the
momentum-conserving delta function), and the energies of
antineutrinos, electrons, and protons can be used as new
integration variables, yielding
\begin{eqnarray}
Q_{\bar\nu}=\frac{4Be}{(2\pi)^5\hbar^9c^6}
\sum_{n,l,\sigma_p,\sigma_n,\gamma}\int_0^\infty dE_{\bar\nu}
\int_0^\infty dp_n \int_{a_e}^\infty dE_e\int_{a_p}^\infty dE_p
\frac{p_n E_{\bar\nu}^3 E_e E_p f_n(1-f_p)(1-f_e)\delta (\Delta
E)\Theta (u)M}{\sqrt{E_{p}^2-m_p^2c^4 - 2Be\hbar cm}
\sqrt{E_{e}^2-m_e^2c^4 - 2Be\hbar cn} }, \label{eq:gfE}
\end{eqnarray}
where $a_e = \sqrt{m_e^2c^4 + 2Be \hbar cn}$ and $a_p =
\sqrt{m_p^2c^4 + 2Be \hbar cm}$, and $\Theta (u)$ is the usual
Heaviside step function. The values of the new variable
$\gamma=\pm 1$ label the cases in which $p_{ez}$ and $p_{pz}$ have
equal and opposite signs, respectively, affecting the value of
$u = {\left[p_n^2c^2 - \left(\sqrt{E_e^2 - m^2_ec^4 -2Be\hbar cn}
+\gamma\sqrt{E_p^2 - m^2_pc^4 - 2Be\hbar cm} \right)^2\right]
/(2Be\hbar c)}.$

If the particles are in chemical equilibrium, the Fermi-Dirac
factors suppress those reactions in which the difference between
the energy of the reacting particles and their corresponding Fermi
energies is much larger than $kT$. This fact allows to also change
the integration variable for the neutrons, as $p_ndp_n=m_n^*dE_n$,
and to set $E_i = E_{Fi}$ in those terms where a variation of the
energy of order $kT$ makes very little difference.

For instance, in the Appendix it is shown that there is a range in
which $F_{m,n}(u)$ behaves as an oscillating function and that,
outside this range, this function decays exponentially. The
average oscillation period is $\approx 4
\sqrt{{\max\{n,m\}}/{\min\{n,m\}}}> 1,$ much larger than the
variation in $u$ due to an energy change $\sim kT$, $\delta u\sim
kT/\Delta E_p\ll 1$. Therefore, in the variable $u$, we can safely
set $E_i = E_{Fi}$, and obtain
$u = r - \left(\sqrt{n_M - n}+\gamma\sqrt{m_M - m}\right)^2$, where $r$, $n_M$ and $m_M$
were defined in (4).

With the factors $\sqrt{E_e^2 - m_e^2 - 2Be\hbar cn}$ and
$\sqrt{E_p^2 - m_p^2 - 2Be\hbar cm}$ in the denominator of eq.
(\ref{eq:gfE}), we {\it cannot} always do the same. Since they
could eventually vanish for the highest Landau band of each
particle, a variation of the energy of order $kT$ could become
crucial in the emission rate we want to calculate.

Let us first consider a situation in which this is not an issue,
i.e., none of the Landau levels of the particles is too close to
the respective Fermi energy,
\begin{equation}
\left|E_{Fe} - \sqrt{m_e^2c^4 + 2Be\hbar cn}\right| \gg kT
\textrm{ } \textrm{ and  } \textrm{ }\left|E_{Fp} - \sqrt{m_p^2c^4
+ 2Be\hbar c m}\right| \gg kT \label{eq:cond}
\end{equation}
\noindent for all $n$ and $m$. In this case, we can set $E_e =
E_{Fe}$ and $E_p = E_{Fp}$, obtaining
\begin{equation}
\overline{Q}_{\bar\nu} = \frac{457 \pi m_n^* m_p
E_{Fe}(kT)^6}{80640 \hbar^{10}c^5} \sum_{\gamma=\pm 1}
\sum_{\sigma_n=\pm 1} \sum_{\sigma_p=\pm 1} \sum_{n=0}^{\lfloor
n_M \rfloor}\sum_{l=0}^{l_{max}}\frac{\Theta(u)M}{\sqrt{n_M -
n}\sqrt{m_M - m}} \textrm{ , } \label{eq:nogen}
\end{equation}
where the symbol $\lfloor \textrm{ } \rfloor$ denotes the integer
part, and $l_{max}={\lfloor m_M-{1\over 2}+{\sigma_p g_p\over
4}\rfloor}$.

\begin{figure}
\centering\includegraphics[width=7cm]{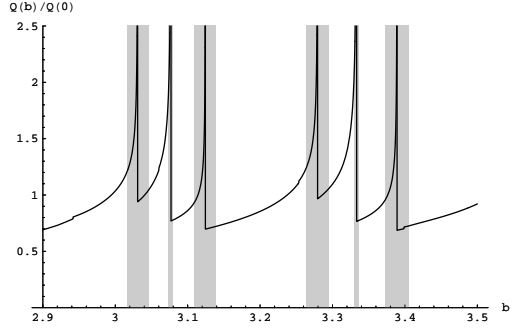}
\caption{Magnetic-field dependent emissivity (normalized by the
zero-field value) at low temperatures, as approximated by eq.
(\ref{eq:nogen}), for $r=100/b$ and $n_M=m_M=40/b$, corresponding
to constant particle number densities, $\rho_e=\rho_p=0.25\rho_n$,
and a magnetic field $B=6.8\times 10^{16}b\left(\rho_n/10^{38}{\rm
cm}^{-3}\right)^{2/3}$ G. The shading identifies those regions
where thermal effects at temperatures $T\sim 10^9$ K would
strongly affect the emissivity.} \label{fg:sum}
\end{figure}

In Fig.~\ref{fg:sum}, we show a numerical calculation of this
double sum for fixed densities of all particles, and a variable
magnetic field. The true emissivity will deviate from this result
(and the formal divergences will be smoothed, as discussed below)
in those regions where the condition (\ref{eq:cond}) does not
hold. As an example, the shading in our graph identifies these
regions for a temperature $\sim 10^9$ K. Still, it is instructive
to note that the peaks come in ``triplets'', in which the central
(and strongest) peak corresponds to a spin-degenerate electron
Landau level coinciding with its Fermi energy, whereas the two
flanking peaks correspond to similar coincidences for the
(non-degenerate) proton Landau levels, for opposite spins. Their
symmetric spacing arises automatically when the proton and
electron densities are equal, as is the case in charge neutrality,
when protons and electrons are the only charge carriers, but does
not necessarily occur when other charge carriers (muons, mesons,
or hyperons) are also present in the system.

Much less conspicuous are the discontinuities caused by the step
function $\Theta(u)$ that ensures momentum conservation along the
magnetic field. Since, as $u\to 0$, the Laguerre functions behave
as $F_{n,l}\sim u^{|n-l|}$ (see eq. \ref{eq:a1}), a true
discontinuity in the emissivity (rather than in one of its
derivatives) arises only for terms with $n=l$ (in which the
spatial wave functions of protons and electrons coincide). Some of
these are barely visible in Fig.~\ref{fg:sum}, at $b\approx$ 2.94,
3.06, 3.25, and 3.40. These will also tend to be washed out by
thermal effects, so we doubt that they could have any
astrophysical significance.

An average value for $\overline{Q}_{\bar\nu}$ can be obtained by
making some approximations for the behavior of the Laguerre
functions $F_{n,l}$ when $n$, $l$, and $|n-l|$ are all large. We
take the average expression for the functions $F_{m,n}^2(u)$ in
the interval where they oscillate (eq. [\ref{eq:a13}]), and
$F_{m,n}^2(u) = 0$ where they undergo an exponential decay (see
Appendix). Then, approximating the sums over $n$ and $l$ as
integrals and neglecting the spin dependence of the particles'
energies, we find
\begin{equation}
\overline{Q}_{\bar\nu} = G_F^2 (1+3g_A^2)\frac{457\pi
m_n^*m_pE_{Fe}(kT)^6}{20160 \textrm{ } \hbar^{10}c^5},
\label{eq:Bo}
\end{equation}
which agrees with the unmagnetized case (Lattimer et al. 1991).

The approximations leading up to eq. (\ref{eq:nogen}) are exact in
the limit of vanishing temperature, but clearly break down at
finite temperature when one of the Landau levels approaches the
Fermi energy of the respective particle species, so the condition
(\ref{eq:cond}) is not satisfied for this level. The contribution
of the latter to the total emissivity becomes very large (formally
divergent, in the approximation considered above) and is no longer
well-approximated by the corresponding term in the expression
(\ref{eq:nogen}). In order to determine the importance of this
effect, we must compute the contribution of this Landau band
independently and, then, compare it to $\overline{Q}_{\bar\nu}$ as
given by eq. (\ref{eq:Bo}), which still gives a reasonable
approximation to the contribution of all the other bands.

We focus on the case in which one of the Landau levels of the
electrons, labelled as $\hat n$ (either $\hat n=\lfloor
n_M\rfloor$ or $\hat n=\lfloor n_M\rfloor+1$), is close to their
Fermi energy, i. e., $|E_{Fe} - \sqrt{m_e^2c^4 + 2Be \hbar c \hat
n}|\sim kT$, but the same does {\it not} happen for the protons.
The contribution of the electron Landau band $\hat n$ to the
emissivity, obtained from the appropriate term in eq.
(\ref{eq:gfE}), summed over all proton Landau bands,
and using the same approximations for $F_{\hat n,l}^2(u)$ as in
equation (\ref{eq:Bo}), is
\begin{equation} \hat Q_{\bar\nu} = G_F^2(1+g_A^2)
\frac{Be\hbar cm_n^*m_pE_{Fe}}{2^3\pi^5 \hbar^{10}c^5}(kT)^5
\sqrt{\frac{2kT}{E_{Fe}(E_{Fe}^2-m_e^2 c^4)}} I(x), \label{eq:hlr}
\end{equation}
where $x \equiv (E_{Fe} - \sqrt{m_e^2c^4 + 2Be \hbar c \hat
n})/(kT)$ and
\begin{equation}
I(x) \equiv \int_0^{\infty}\int_0^\infty \frac{t^3(t+u-x)du
dt}{u^\frac{1}{2}(e^{x-u}+1)(e^{t+u-x}-1)}. \label{eq:Ix}
\end{equation}

\begin{figure}
\centering\includegraphics[width=7cm]{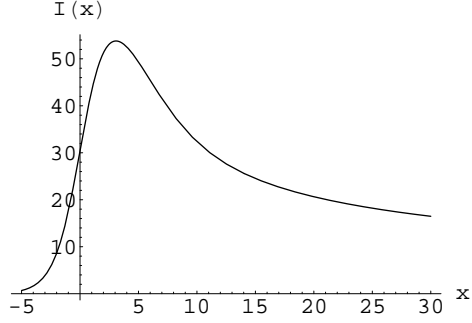} \caption{The
function $I(x)$, evaluated numerically from eq. (\ref{eq:Ix}). }
\label{fg:integral}
\end{figure}

\noindent In Figure 2, we can see that this function has a peak at
$x=3.08$, where its value is $53.8$, decreasing as $I(x) \approx
457{\pi}^6/(5040\sqrt{x}) $ for $x\gg 3$, and exponentially for
$x<0$. The smooth shape of this function (left-right reversed,
since $x$ is a decreasing function of $B$) is the true shape of
the divergent peaks of Fig. 1 at finite temperature.

\section{Discussion}
\label{sec:discussion}

We compare the emissivity contribution of the highest
significantly populated electron Landau band to the unmagnetized
emissivity (which approximates the contribution from all the other
Landau bands) by computing the ratio
\begin{equation}
\frac{\hat Q_{\bar\nu}}{\overline{Q}_{\bar\nu}}\approx
\frac{I(x)}{123} \frac{Be \hbar c}{\sqrt{kT E_{Fe}(E_{Fe}^2 -
m_e^2c^4)}}\approx{I(x)\over 123}{\Delta E_e\over\sqrt{kTE_{Fe}}}.
\label{eq:ratio}
\end{equation}
As the density is increased or the magnetic field is decreased,
the emissivity oscillates as new Landau bands start populating.
Assuming, for definiteness, that the control parameter is
$E_{Fe}$, and ignoring variations of $E_{Fp}$ for now, these
variations are roughly periodic with a period $\Delta E_e$, on
which asymmetric peaks occur (sharply rising as the Landau level
starts populating, slowly decreasing afterwards), with a total
width at half maximum $\sim 10kT$. If $\Delta E_e\gg 10kT$, the
fractional amplitude of the oscillations is $\sim 0.4\Delta
E_e/\sqrt{kTE_{Fe}}$, becoming comparable to the total emissivity
when the level spacing is the geometric mean of $kT$ and the Fermi
energy or, equivalently, $B\sim 10^{17}\sqrt{T_8\rho_{37}}{\rm
G}$, probably higher than is reached in magnetars.

In the case of the protons, the situation is different.
Since they interact strongly, their dispersion relation
is not exactly the one shown in eq. (\ref{eq:energies}).
Strong interactions, mainly with the neutrons, broaden the Landau levels of the protons.
If this broadening is of order or larger than the spacing between the levels,
the effect of emissivity oscillations  given by the
filling of their consecutive Landau bands tends to disappear.
On the other hand, even  if the protons did not interact strongly, the importance of the
emissivity contribution of their highest populated Landau band would be
subdominant in comparison to the one of the electrons. This can be seen directly from
eq. (\ref{eq:ratio}), taking into account two facts. First, proton Landau
bands are not spin degenerate, so the expression for the ratio
$\hat Q_{\bar\nu}/\overline{Q}_{\bar\nu}$  is reduced by a factor 1/2.
Second, their larger mass implies (at the same density) a smaller
Landau-level spacing, yielding a more stringent condition to see
substantial oscillations, and a smaller amplitude for these. It
also implies that the last expression in eq. (\ref{eq:ratio}) would
not be directly applicable, but would have to be replaced by a similar one, in
which the mass term $m_pc^2$ is subtracted from the Fermi energy.


In a neutron star in diffusive equilibrium, the redshifted total
chemical potential (including electrostatic potential energy) for
each particle species is uniform throughout the region where the
respective species is present. Since the redshift factor is a
smooth function of position, so will be the local (or
``intrinsic'') chemical potentials, i. e., the Fermi energies
considered in this work except at phase boundaries, where the
electrostatic potentials can be discontinuous. We stress that this does {\it not} imply
that the particle densities are smooth functions of position, but
will rather have oscillations superimposed on the smooth increase
with increasing Fermi energy, as illustrated, e.g., in Fig. 2 of
Dib \& Espinosa (2001).

More directly to the point of this paper, even if the magnetic
field is uniform, the smooth spatial variation of the Fermi
energies will make the emissivity an oscillating function of
position. Shells of larger and smaller emissivity will alternate
within the star. If the magnetic field strength is truly uniform,
these shells will be spherical, constant-density surfaces,
otherwise they can be arbitrarily convolved.

The observational relevance of this is not clear. As said above,
in each ``period'' $\Delta E_e$, the emissivity peak has a width
$\sim 10 kT$ and a fractional height $\sim 0.4\Delta
E_e/\sqrt{kTE_{Fe}}$, so its fractional contribution to the total
emissivity over the corresponding period is $\sim(10kT/\Delta
E_{e})(0.4\Delta E_e/\sqrt{kTE_{Fe}})=4\sqrt{kT/E_{Fe}}$. The
degeneracy condition $kT\ll E_{Fe}$ is amply satisfied in all
neutron stars more than a few seconds old, so the contribution of
the peaks to the total emissivity is very small.

\section{Conclusions}
\label{sec:co}

We have shown that, in the regime in which the Landau level
separation is intermediate between $kT$ and the Fermi energies
(excluding $mc^2$), the direct Urca emissivity is strongly
enhanced when the Fermi energy of electrons or protons lies $\sim
3 kT$ above the bottom of any of the Landau bands. Thus, it will
oscillate as a function of density or field strength. The
oscillation amplitude becomes comparable to the total emissivity
when the level separation is near the geometric mean of the other
two energy scales. The required conditions may be present in
magnetars, but this is not guaranteed. Even if it were the case,
the large number of emissivity oscillations expected inside a
given star could average out to give nearly the same global
cooling rate as in an unmagnetized star, so their observational
consequences are far from obvious.

\appendix

\section{Some properties of the Laguerre functions}
\label{sec:ap}

For easy reference, in this Appendix we summarize some useful properties of the
Laguerre functions, $F_{m,n}(u)$.
These results were obtained by Kaminker $\&$ Yakovlev
(1980, 1981).

The Laguerre functions with indices $m,n=0,1,2, ...$ are defined
as
\begin{eqnarray} F_{m,n}(u) = \left\{ \begin{array}{cc}
\sqrt{n!/m!} (-1)^{a}e^{-\frac{u}{2}}u^\frac{a}{2}L_n^{a}(u),\quad
& {\rm for}\quad m\geq n \\
(-1)^a F_{n,m}(u) & {\rm otherwise}
\end{array} \right.
\label{eq:a1}
\end{eqnarray}
where $a = m-n$ and $L_n^{a}(u)$ correspond to the Laguerre polynomials,
\begin{equation}
L_n^{a}(u) = \sum_{k = 0}^n \frac{\Gamma(n+a+1)}{\Gamma(k+a+1)}
\frac{(-u)^k}{k! (n-k)!}. \label{eq:a2}
\end{equation}
Defining $f(u) = \frac{1}{4u^2}(u-u_-)(u-u_+)$, where $u_\pm = (m+n+1) \pm \sqrt{4mn + 2(m+n)+2}$,  approximate expressions for $F_{m,n}(u)$  for different ranges of $u$  in the limit $n,m, |n -
m|\gg 1$, can be given as follows:
\begin{itemize}
\item For
$u_--u\gg\left(\frac{u_-^2}{u_+-u_-}\right)^\frac{1}{3}$,
\begin{equation}
F_{m,n}(u)=(-1)^{\frac{a+|a|}{2}} \left\{ 4\pi u \sqrt{f(u)}
\right\}^{-\frac{1}{2}}e^{u\sqrt{f(u)}-\phi}. \label{eq:a6}
\end{equation}
\item For
$u-u_+\gg\left(\frac{u_+^2}{u_+-u_-}\right)^\frac{1}{3}$,
\begin{equation}
F_{m,n}(u)=(-1)^m \left\{ 4\pi u \sqrt{f(u)}
\right\}^{-\frac{1}{2}}e^{-u\sqrt{f(u)}-\phi}, \label{eq:a6a}
\end{equation} where
\begin{equation}
 \phi =
\frac{u_- +
u_+}{4}\ln\left(\frac{\left\{\sqrt{|u-u_+|}-\sqrt{|u-u_-|}\right\}^2}{u_+-u_-}\right)\textrm{
}+\textrm{ }
\frac{\sqrt{u_-u_+}}{2}\ln\left(\frac{\left\{\sqrt{u_-|u-u_+|}+\sqrt{u_+|u-u_-|}\right\}^2}{u\{u_+-u_-\}}\right).
\label{eq:a7}
\end{equation}
\item For $| u - u_- | \ll u_-,u_+-u_-$,
\begin{equation}
F_{m,n}(u) =
(-1)^{\frac{a+|a|}{2}}\left\{\frac{4}{u_-(u_+-u_-)}\right\}^{\frac{1}{6}}
{\rm
Ai}\left(\{u_--u\}\left\{\frac{u_+-u_-}{4u_-^2}\right\}^\frac{1}{3}\right).
\label{eq:a8}
\end{equation}
\item For $| u - u_+ | \ll u_+ -u_-$,
\begin{equation}
F_{m,n}(u) =
(-1)^{m}\left\{\frac{4}{u_+(u_+-u_-)}\right\}^{\frac{1}{6}} {\rm
Ai}\left(\{u-u_+\}\left\{\frac{u_+-u_-}{4u_+^2}\right\}^\frac{1}{3}\right),
\label{eq:a81}
\end{equation}
where ${\rm Ai}(x)$ corresponds to the Airy function of the first
kind (see, for example, Lebedev 1972).The asymptotic behavior of
this function,
\begin{eqnarray}
{ {\rm Ai}(x) \approx \left\{ \begin{array}{cc}
\frac{\pi^{-\frac{1}{2}}}{2}x^{-\frac{1}{4}}e^{-\frac{2}{3}x^\frac{3}{2}}
& x \rightarrow \infty \\
\pi^{-\frac{1}{2}}|x|^{-\frac{1}{4}}\cos(\frac{2}{3}|x|^\frac{3}{2}-
\frac{\pi}{4} ) & x \rightarrow -\infty \textrm{,}
\end{array} \right.  }
\end{eqnarray}
shows us that the Laguerre functions oscillate between $u_-$ and
$u_+$ and decay exponentially outside this range.

\item For
$u-u_- \gg \left( \frac{u_-^2}{u_+-u_-}\right)^{\frac{1}{3}}$ and
$u_+-u \gg \left( \frac{u_+^2}{u_+-u_-}\right)^{\frac{1}{3}}$, we
have
\begin{equation}
F_{m,n}=\left\{ \pi u \sqrt{-f(u)} \right\}^{-\frac{1}{2}}
\cos\left(u\sqrt{-f(u)} - \varphi \right), \label{eq:a9}
\end{equation}
where
\begin{equation}
\varphi = \frac{u_- +
u_+}{4}\sin^{-1}\left(\frac{u_-+u_+-2u}{u_+-u_-}\right)\textrm{ }
-\textrm{ }\frac{\sqrt{u_+u_-}}{2}\sin^{-1}\left(\frac{2u_-u_+ -
u\{u_-+u_+\}}{u\{u_+-u_-\}}\right)-\frac{\pi}{4}\left(
m+n+2a+|a|\right). \label{eq:a10}
\end{equation}
In this work we use the average expression for $F_{m,n}^2(u)$ in this
range,
\begin{equation}
\langle F_{m,n}^2\rangle = \pi^{-1}\left\{
(u-u_-)(u_+-u)\right\}^{-\frac{1}{2}}. \label{eq:a13}
\end{equation}
\end{itemize}

\begin{acknowledgements}

We thank D. G. Yakovlev for providing us with the unpublished
manuscript by Kaminker \& Yakovlev (1980). We also acknowledge
support from a CONICYT Doctoral Fellowship and from FONDECYT
Regular Grants \# 1020840, 1030363, and 1030254.

\end{acknowledgements}

\end{document}